\documentclass[aps,prb,twocolumn,groupedaddress]{revtex4-1}

\usepackage{graphicx}
\usepackage{dcolumn}
\usepackage{bm}
\usepackage{color}

\begin{document}


\title{Single crystal growth and physical properties of a new uranium compound URhIn$_5$ }


\author{Yuji Matsumoto$^{1,2}$\thanks{E-mail address: matsumoto.yuji@nitech.ac.jp},  Yoshinori Haga$^{1}$, Naoyuki Tateiwa$^{1}$, Hironori Sakai$^{1}$, Tatsuma D. Matsuda$^{1}$, 
\\Etsuji Yamamoto$^{1}$,   and Zachary Fisk$^{1,3}$}
\affiliation{$^{1}$Advanced Science Research Center, Japan Atomic Energy Agency, Tokai, Ibaraki 319-1195, Japan\\
$^2$Graduate School of Engineering, Nagoya Institute of Technology, Nagoya, Aichi 466-8555, Japan\\
$^{3}$University of California, Irvine, CA 92697, USA}


\date{\today}

\begin{abstract}
We have grown the new uranium compound URhIn$_5$ with the tetragonal HoCoGa$_5$-type by the In self flux method.
In contrast  to the nonmagnetic ground state of the isoelectronic analogue URhGa$_5$, URhIn$_5$ is an antiferromagnet with antiferromagnetic transition temperature $T_{\rm N}$ = 98 K.
The moderately large electronic specific heat coefficient $\gamma$ = 50 mJ/K$^2$mol demonstrates the contribution of 5$f$ electrons to the conduction band.
On the other hand, magnetic susceptibility in the paramagnetic state roughly follows a Curie-Weiss law with a paramagnetic effective moment corresponding to a localized uranium ion.
The crossover from  localized  to itinerant character at low temperature may occur around the characteristic temperature 150 K where the magnetic susceptibility and electrical resistivity show a marked anomaly.

\end{abstract}


\maketitle

\section{Introduction}

Actinide elements and their compounds are characterized by the 5$f$ electrons.  
Because of the large spatial extent of the wave functions, 5$f$ electrons are sensitive to the physical or chemical environments surrounding them.
Some of the most astonishing behavior can be found in plutonium metal.
It shows successive structural phase transitions with 6 different crystal structures as a function of temperature.\cite{Ott87}
Among these structures, cubic $\delta$-phase plutonium has about 20 \% larger unit cell volume than that of the $\alpha$-phase at room temperature.
Such a large volume change might accompany a change in the valence state.
A magnetic ground state is theoretically predicted for $\delta$-Pu with its large volume, although no magnetic moment has  been detected experimentally.\cite{heffner06,soderlind04}
Valence change now  attracts attention because of the possibility of a novel superconducting pairing interaction mediated by valence fluctuations.\cite{miyake02}
Recent experimental discovery of the isostructural superconductors PuTX$_5$ (T = Co and Rh, X = Ga and In) with the tetragonal HoCoGa$_5$-type structure has stimulated those discussions.\cite{Sarrao02,Wastin03,bauer12a,bauer12b}
Actinide elements and their compounds have accordingly been extensively studied despite their attendant experimental difficulties.

In this paper, we report an attempt to change the ground state of a 5$f$ electron by changing its relative volume.
In this context, compounds with the tetragonal HoCoGa$_5$-type (115) structure are suitable because a series of compounds with the same structure widely exist in rare earth and actinide system. 
Due to the occurrence of unconventional superconductivity, the heavy Fermion state, and magnetic ordering and their coexistence  in Ce and Pu analogues, physical properties have been extensively investigated in this system and as a consequence well characterized high-quality single crystals are available.\cite{Joe12}

The crystal structure of the so-called 115 compounds is shown in Fig. 1. 
The uniaxially distorted  AX$_3$-layers with similar arrangement to the cubic AuCu$_3$-type structure and TX$_2$ layers are stacked sequentially along the [001] direction.  
The 115 structure is generally written as ATX$_5$ where A is a rare earth, actinide or tetravalent transition metal.  Group 9 - 11 transition metals can be a constituent of the actinide 115 compounds, but only group 9 transition metals are found forming the rare earth 115s.  Because of its larger ionic radius, In occupies the X site in the 115 compounds with larger unit cell volume, namely A = rare earths in most cases, while uranium with its small ionic radius forms 115 structures stable only when Ga occupies the X site\cite{Moreno03,Ikeda05b}.

Consistent with this, the  Fermi surfaces of the uranium 115 compounds experimentally determined by means of de Haas-van Alphen (dHvA) or angular-resolved photoemission (ARPES) are well explained by band calculations where 5$f$ electrons are treated as band electrons, namely 5$f$ electrons can be regarded as itinerant.\cite{Tokiwa01,Tokiwa01b,Tokiwa02,Tokiwa02b,Ikeda02,Ikeda05,Fujimori2006}
Among them, URhGa$_5$ is a nonmagnetic metal with semimetallic behavior.\cite{Ikeda02}  Its magnetic susceptibility is almost constant as a function of temperature, in contrast to the magnetic ground state observed in NpRhGa$_5$\cite{Aoki2005JPSJ,Colineau2005} and the temperature dependent paramagnetic behavior in the superconductor PuRhGa$_5$.\cite{Haga2005JPSJ}  The small carrier number determined from dHvA experiments and the small electronic specific heat are consistent with the 5$f$-itinerant band calculations demonstrating the small overlap of the conduction bands and resulting small density of states at the Fermi energy.\cite{Maehira2006NJP}

We report here the first synthesis of a single crystalline URhIn$_5$ with a significantly larger unit cell volume than URhGa$_5$.  As might be expected, a magnetic ground state is realized in this compound.

\section{Experimental}

\begin{figure}
\begin{center}
\includegraphics[width=8cm]{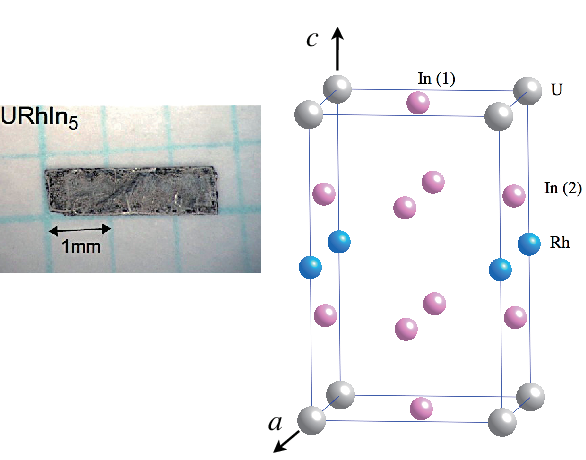}
\end{center}
\caption{Photograph of a single crystal of URhIn$_5$ and its crystal structure. }
\label{crystal}
\end{figure}

Single crystals of URhIn$_5$ were grown using the In self flux method in a sealed quartz tube, using 3N8 (99.98 \% purity) U,  4N Rh and 5N In. 
A pre-reacted U-Rh-In alloy and excess indium  were placed in an alumina crucible and sealed in an evacuated quartz tube. 
The sealed tubes were heated to 1150 $^{\circ}$C, soaked for 12 h, then cooled down to 150 $^{\circ}$C in 168 h. 
The excess In was spun off in a centrifuge.
A photograph of single crystal URhIn$_5$ is shown in Fig \ref{crystal}. 
We also attempted unsuccessfully growths of UCoIn$_5$ and UIrIn$_5$. 

The chemical composition and homogeneity of the crystals were confirmed by electron-probe micro analysis (EPMA) using JEOL JX-8900. 
The crystal structure  of URhIn$_5$  was determined via single crystal X-ray  diffraction measurements with Mo K$_{\rm \alpha}$ radiation using an R-AXIS RAPID (Rigaku) diffractometer.  Of the 1811 Bragg reflections collected, 175 were unique.  Crystallographic parameters were determined and refined using the SHELX-97 program. \cite{sheldrick97}
Both EPMA and X-ray diffraction suggested that there is no significant deviation from the stoichiometry.

Table 1 shows atomic coordinates and the equivalent isotropic atomic displacement parameters $B_{\rm eq}$ of URhIn$_5$ at 292 K.
The lattice constants $a$ and $c$ of URhIn$_5$ are 4.6205 and  7.4168 ${\rm \AA}$, respectively, and the positional parameter of the In(2) site was determined as (0 0 0.3018).  As mentioned in the Introduction, the local atomic arrangement around the uranium site is similar to that in the cubic AuCu$_3$-type structure.  Using the crystallographic parameters determined in this study, the bond angle between U-In(2)-U was determined as 91.8$^\circ$ being slightly larger than the 90$^\circ$ expected for  the ideal AuCu$_3$ structure, indicating a small tetragonal distortion.  Compared to the existing cubic UIn$_3$ ($a$ = 4.701 ${\rm \AA}$)\cite{Begun1993}, the  'UIn$_3$' building block in URhIn$_5$ is compressed 1.9 \% and 4.7 \% along $a$ and $c$ directions, respectively.

Specific heat measurements were  performed using a home-built apparatus by the heat relaxation method in a $^3$He cryostat.
The electrical resistivity measurements were performed using a conventional DC four-terminal technique.
DC magnetization measurements were performed using a superconducting quantum interference device magnetometer (Quantum Design MPMS) from 300 to 2 K in magnetic fields up to 7 T. 

To study the effect of pressure on the antiferromagnetic transition temperature ${T_{\rm N}}$ in URhIn$_5$, we measured the temperature dependence of the electrical resistivity using a piston-cylinder type high pressure cell with Daphne 7373 oil as the pressure-transmitting medium\cite{tateiwa}.

\begin{figure}
\begin{center}
\includegraphics[width=8cm]{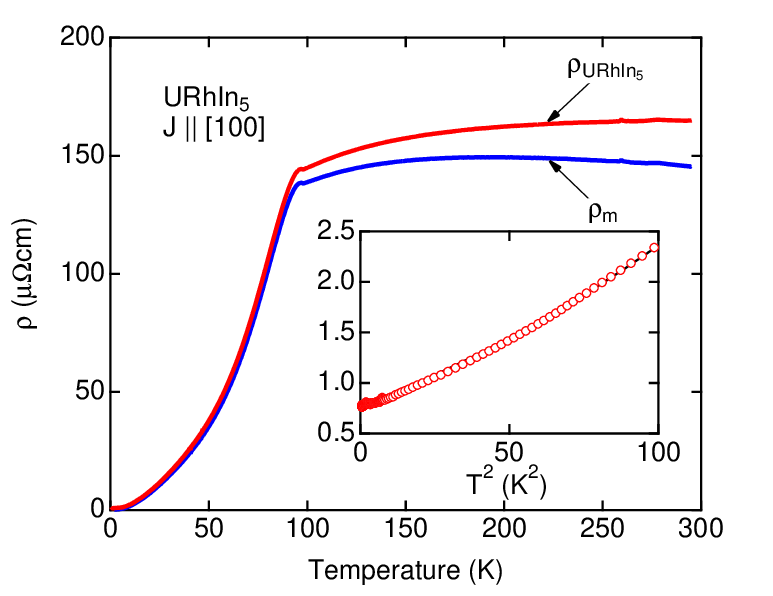}
\end{center}
\caption{Temperature dependence of the resistivity $\rho_{\mathrm{URhIn_5}}$ and magnetic resistivity  $\rho_\mathrm{m}$ = $\rho_{\mathrm{URhIn_5}}$ - $\rho_{\mathrm{ThRhIn_5}}$  of URhIn$_5$ for current along [100] direction.  $\rho_{\mathrm{ThRhIn_5}}$ is taken from ref. \cite{TD07}.  }
\label{rho}
\end{figure}

\begin{table}
\caption{Atomic coordinates and equivalent isotropic atomic displacement parameters $B_{\mathrm{eq}}$ of URhIn$_5$ at 292 K.  The number in parenthesis is the standard deviation of the last digit.  The least-squares refinement was based on 175 independent reflections and converged with a conventional agreement factor $R_1$ = 0.0319.}
\label{xray}
\begin{center}
\begin{tabular}{ccccc}
\hline
\multicolumn{1}{c}{Atom} & \multicolumn{1}{c}{x}  & \multicolumn{1}{c}{y}  &  \multicolumn{1}{c}{z} &  \multicolumn{1}{c}{$B_{\mathrm{eq}}$}   \\
\hline
U  &  0   & 0  &  0 &  0.64(3) \\
Rh  & 0  & 0   & 1/2   &  0.71(4) \\
In(1)  &   1/2   &   1/2    &   0    &  0.87(3) \\
In(2) & 1/2   & 0   & 0.3018(1)   &  0.85(3)  \\
\hline
\end{tabular}
\end{center}
\end{table}

\begin{figure}
\begin{center}
\includegraphics[width=8cm]{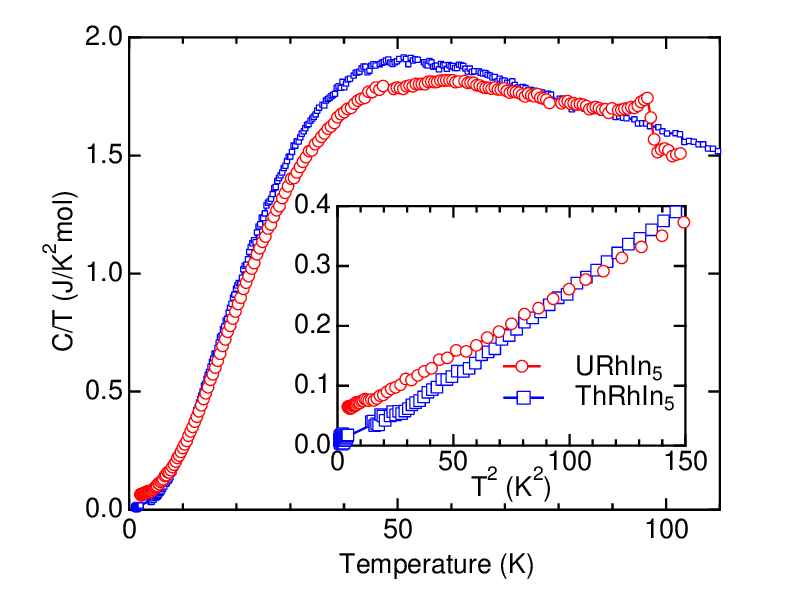}
\end{center}
\caption{Temperature dependence of the C/T of URhIn$_5$ and ThRhIn$_5$. The open circle indicates the temperature dependence of URhIn$_5$. The open square indicates the temperature dependence of ThRhIn$_5$. }
\label{c}
\end{figure}

\section{Results}

We measured the temperature dependence of the resistivity, magnetic susceptibility, specific heat and lattice parameter of URhIn$_5$. 

First, we show the temperature dependence of resistivity in URhIn$_5$ for current along the [100] direction in Fig. \ref{rho}.  
Also plotted in Fig. \ref{rho} is the magnetic contribution to resistivity where the resistivity of the non-magnetic analogue ThRhIn$_5$ for the lattice contribution was subtracted:  $\rho_m$  ( = $\rho_{\mathrm{URhIn_5}}$ - $\rho_{\mathrm{ThRhIn_5}}$ ). \cite{TD07}
	The $\rho_m$ increases with decreasing temperature, suggesting an enhancement of magnetic scattering at lower temperatures.
	It has a maximum around 170 K.  The gradual increase of magnetic resistivity is frequently seen in other uranium or rare earth compounds and is attributed to the Kondo effect. This point will be discussed later.
With further decreasing temperature,  $\rho_m$ has a weak kink at 98 K and a subsequent hump, implying a phase transition with formation of a gap at the Fermi surface.  A distinct anomaly in magnetic susceptibility at 98 K, without spontaneous moment as described below, strongly suggests that an antiferromagnetic transition takes place at $T_N$ =  98 K.

The $\rho_m$ rapidly decreases with decreasing temperature below $T_{\rm N}$. 
At low temperatures, the electrical resistivity can be well fit with the expression 

$\rho_m(T) = \rho_0 + AT^2 + DT/\Delta (1+2T/\Delta)e^{-\Delta/T}$, 

where $\rho_0$ is the residual resistivity,  $A$ is the coefficient of $T^2$ term arising from electron-electron scattering in Fermi liquid, $D$ involves the electron-magnon and the spin-disorder scattering and $\Delta$ is the magnitude of the energy gap in magnetic excitation.  
By fitting this equation to the experimental data, we obtained  $\rho_0$ = 0.78 $\mu \Omega$cm , $A$ = 0.013 $\mu \Omega$cm/K$^2$, $D$ = 0.023 $\mu \Omega$cm/K  and $\Delta$ = 37 K.  The large residual resistivity ratio RRR  = 220 demonstrates the extremely high quality of the present single crystal sample.

Figure \ref{c} displays the temperature dependence of specific heat $C(T)$ divided by temperature for URhIn$_5$ and ThRhIn$_5$.  
A $\lambda$-shaped anomaly corresponding to a second order phase transition is observed at 98 K. 
At low temperature, specific heat involves electronic specific heat and phonon contributions $C = \gamma T + \beta T^3$, where $\beta$ is related to the Debye temperature.  We obtained the electronic specific heat coefficient $\gamma$ and Debye temperature $\theta_{\rm D}$ of URhIn$_5$ as 50 mJ/mol K$^2$ and 187 K, respectively, and  those of ThRhIn$_5$ as 10 mJ/mol K$^2$ and 179 K, respectively.  
The effective mass of URhIn$_5$ is weakly enhanced.
The specific heat of reference compound ThRhIn$_5$ is larger than that of URhIn$_5$ at high temperature due to the differing  lattice specific heat.
Therefore, we could not obtain the magnetic specific heat $C_m$ of URhIn$_5$ by subtracting the specific heat of ThRhIn$_5$.

\begin{figure}
\begin{center}
\includegraphics[width=8cm]{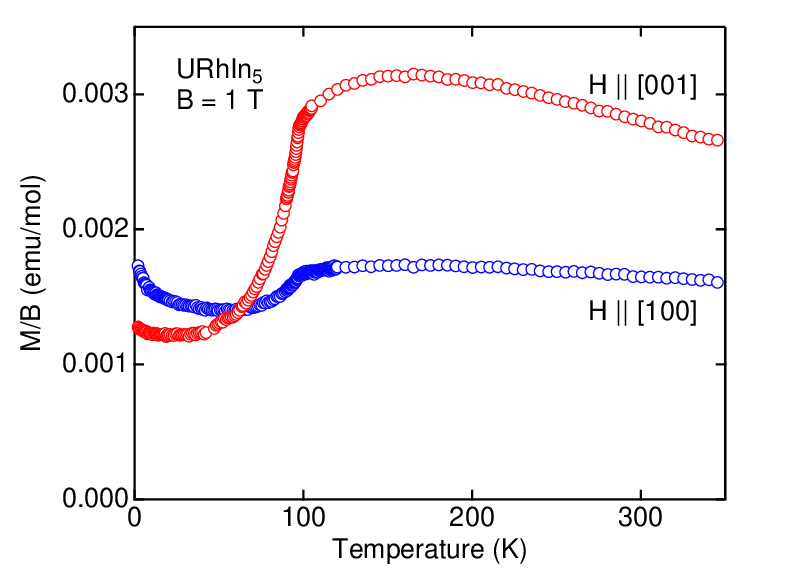}
\end{center}
\caption{Temperature dependences of magnetic susceptibility of URhIn$_5$ at 1 T for a magnetic field along the [001] and [100] directions.  
The open circles indicate the susceptibility for magnetic field along the [001] direction.  
The closed squares indicate the susceptibility for magnetic field along the [100] direciton. }
\label{chi}
\end{figure}

\begin{figure}
\begin{center}
\includegraphics[width=8cm]{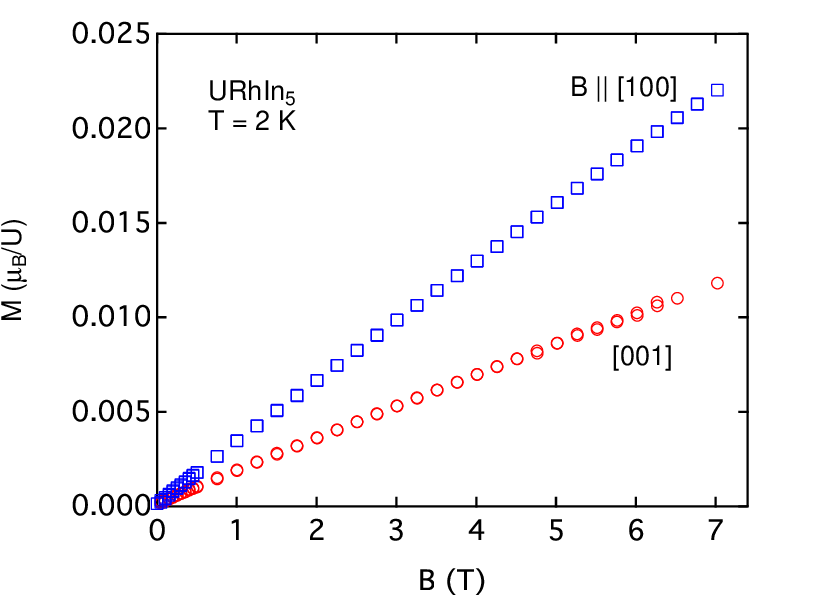}
\end{center}
\caption{Magnetism vs. magnetic field of URhIn$_5$ at 2 K for a magnetic field along [001] and [100] direction.  
The open circles indicate the susceptibility for magnetic field along [001] direction.  
The closed squares indicate the susceptibility for magnetic field along [100] direction. }
\label{mh}
\end{figure}

Figure \ref{chi} shows the magnetic susceptibility of URhIn$_5$ in a magnetic field of 1 T for magnetic field along the [001] and [100] directions. 
The magnetic susceptibility at room temperature has a marked anisotropy with magnetic easy axis along [001].
The susceptibility increases with decreasing temperature and has a maximum around 150 K. 
	The magnitude of the susceptibility is significantly smaller than that expected for a free paramagnetic uranium ion U$^{3+}$ or U$^{4+}$.  The reduction of the magnetic moments may  originate from itinerant characteristics of 5$f$ electrons or a Kondo effect with a very large characteristic temperature.  In either case, a temperature-independent Pauli susceptibility arises at low temperatures.  The present temperature dependence of the total susceptibility can be fit with a generalized Curie-Weiss law : $\chi(T) = \chi_0 + \chi_{CW}(T+\theta_p)$, where $\chi_0$ is the Pauli susceptibility and $\chi_{CW}$ is the Curie-Weiss term with a paramagnetic Curie temperature $\theta_p$, giving $\chi_0 = 2.5 \times 10^{-4} {\rm emu/mol}$ and the effective moment roughly corresponding to that of free uranium ion.  $\theta_p$ is extremely large, -400 K, as we inferred above.  It should be noted, however, that only the measurements at higher temperature are used for the determination of the parameters.   These results will be discussed together with other properties in the next section.

\begin{figure}[]
\includegraphics[width=8.8cm]{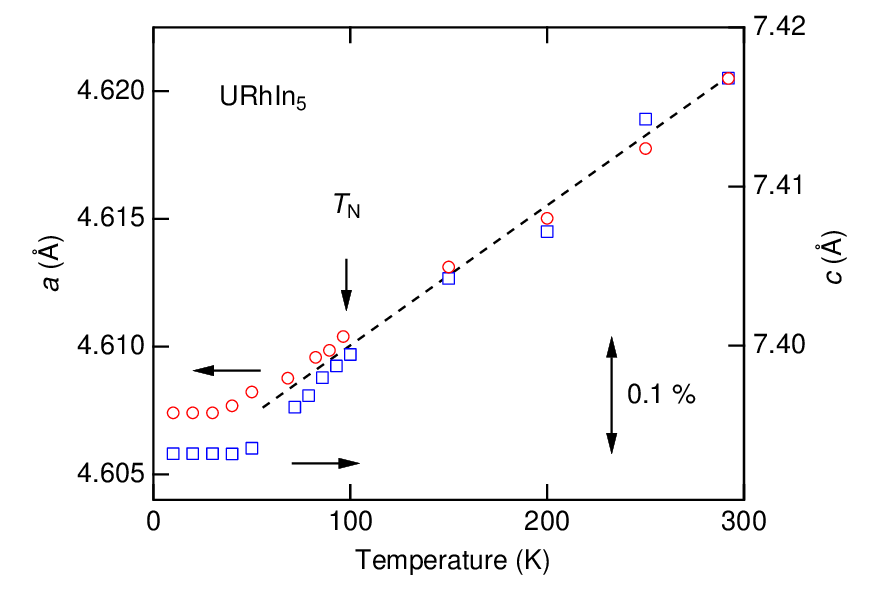}
\caption{\label{fig:epsart}(Color online) Temperature dependence of the lattice parameters}
\end{figure}

The susceptibility rapidly decreases with decreasing temperature below $T_{\rm N}$.
The decreasing of susceptibility at $T_{\rm N}$ for magnetic field along the [001] direction is much larger than that for magnetic field along the [100] direction, indicating that the ordered moments point along the [001] direction in the antiferromagnetic state.  
Recent nuclear quadrupole resonance measurements on URhIn$_5$ agree with this speculation.\cite{sakai13}
The susceptibility slightly increases with decreasing temperature below 40 K.
Such an increase is usually attributed to paramagnetic impurities.  In URhIn$_5$, however, the increase is more significant for the field direction along the [100].  The present increase in magnetic susceptibility might be an intrinsic one and needs further investigation.

Figure \ref{mh} shows the magnetization as a function of the magnetic field for magnetic field along the [001] and [100] directions at 2 K. 
The magnetization  for magnetic field along the [001] and [100] directions increases almost linearly  with magnetic field up to 7 T. 
The magnetization for magnetic field along the [100] direction is about twice  that along the [001] direction.

Here the occurrence of a magnetic ground state is well established.
As mentioned in the Introduction, the larger unit cell volume of URhIn$_5$ as compared to isoelectronic URhGa$_5$ might change the 5$f$ electronic state, leading to a magnetic moment on the uranium site.
It is therefore interesting to investigate the effect of unit cell volume on magnetic ordering.

We show in Fig. 6  preliminary data of the lattice parameters as a function of temperature.  
The antiferromagnetic ordering is seen in the thermal expansion.  
The full scale of the vertical axis in Fig. 6 corresponds to 0.4 \% of length change.
Above $T_{\rm N}$, the lattice parameters linearly decrease with decreasing temperature.
The thermal expansion along the $a$ and $c$ directions is seen to be almost isotropic.
Below $T_{\rm N}$ = 98 K, however, there is a small but  significant difference between the $a$ and $c$ directions.
For the $a$ direction, the slope is almost constant through $T_{\rm N}$, while the $c$ lattice parameter decreases faster below $T_{\rm N}$.
This behavior is significantly different from other magnetically ordered uranium-based 115 compounds.\cite{Kaneko03}
On the other hand, there was no significant temperature variation of the positional parameter of the In(2) site, consistent with the almost isotropic thermal expansion over the entire temperature range.
Using these data, the volume expansion can be deduced (not shown) and will be discussed later.

\begin{figure}[]
\includegraphics[width=7cm]{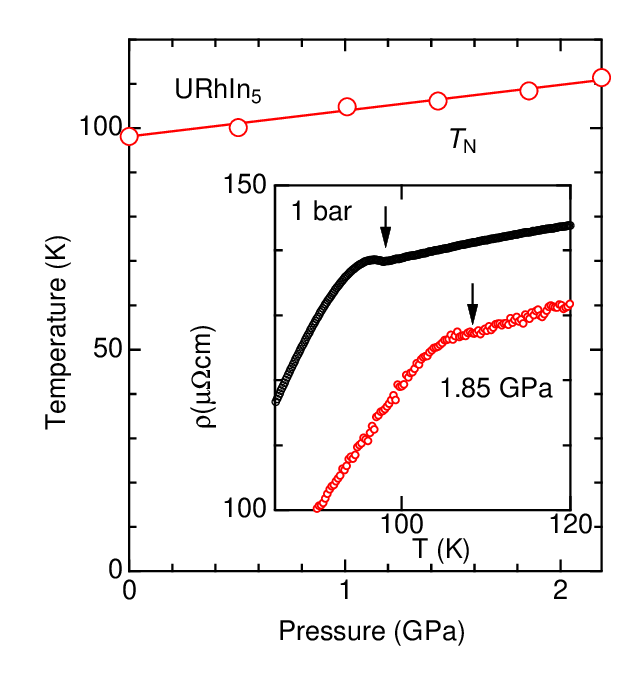}
\caption{\label{fig:epsart}(Color online) Temperature-pressure phase diagram of URhIn$_5$. Inset shows the temperature dependence of the electrical resistivity at 1 bar and 1.85 GPa.}
\end{figure}

 Figure 7 shows the pressure-temperature phase diagram of URhIn$_5$. The inset of the figure shows the temperature dependence of the electrical resistivity $\rho$ at 1 bar and 1.85 GPa near  ${T_{\rm N}}$.  The transition temperature ${T_{\rm N}}$ increases with increasing pressure. The value of ${T_{\rm N}}$ is changed from 98.1 K to 111.4 K at 1 bar and 2.19 GPa, respectively. Assuming the linear pressure dependence of ${T_{\rm N}}$, the value of ${\partial T_{\rm N}}/{\partial P}$ is estimated as 5.8 $\pm$ 0.2 K.

 The antiferromagnetic transition temperature is second order. The initial hydrostatic pressure dependence of $T_{\rm N}$ can be calculated by means of the Ehrenfest relation: $({\partial T_{\rm N}}/{\partial P})_{p{\rightarrow 0}} = {T_{\rm N}}{V_{mol}}({\Delta}{\beta}/{\Delta}{C})$. Here, $V_{mol}$ are the molar volume. ${\Delta}{\beta}$ and ${\Delta}{C}$ are the jump of the thermal expansion coefficient ${\beta} = V{^{-1}}({\partial V}/{\partial T}){_P}$ and the specific heat $C$, respectively, at $T_{\rm N}$. The derived pressure coefficient of $({\partial T_{\rm N}}/{\partial P})_{p{\rightarrow 0}}$ is 5.1 $\pm$ 0.8 K, which is consistent with the experimentally obtained value.

\section{Discussion}

We succeeded in growing a new uranium material URhIn$_5$ and measured its physical properties. 
URhIn$_5$ is supposed to be isoelectronic to URhGa$_5$ with the same crystal structure.
In contrast to the non magnetic semimetallic behavior of URhGa$_5$ the present experimental data suggest that magnetic moments are located on the uranium sites in URhIn$_5$ and order antiferromagnetically with N{\'e}el temperature 98 K.
In this section we discuss the remarkable difference between these compounds.

As mentioned in the Introduction, a uranium atom, in general, has a small atomic radius due to the itinerant tendency of its 5$f$ electrons.  Therefore uranium-based ATX$_5$ compounds so far appear only to be stable when X sites are occupied by Ga.  The unit cell volume of URhIn$_5$ (158.34 ${\rm \AA}$) is 26 \% larger than that of URhGa$_5$ (158.34 ${\rm \AA}$) and hence 5$f$ electrons would have a more localized character.  
In fact,  a modified Curie-Weiss-like behavior with a significant Pauli susceptibility was observed for the temperature dependence of the magnetic susceptibility as shown in the previous section.

The origin of the Pauli susceptibility might be due to an enhanced density states of the conduction bands.
The magnetic part of the electrical resistivity of URhIn$_5$ increases with decreasing temperature above 200 K with the appearance of a broad maximum around 180 K.  
Additionally, the paramagnetic susceptibility follows Curie-Weiss behavior above 200 K, but deviates to have a maximum at 150 K.   
This behavior is reminiscent of the Kondo effect with a characteristic temperature of about 150 K.  
The electronic state below this temperature can be regarded as a renormalized Fermi liquid state where the spin degrees of freedom at high temperature are quenched via the Kondo effect.  
The resulting renormalized effective mass of the Fermi liquid is reflected in an enhanced electronic specific coefficient $\gamma = R \ln 2 / T_{\rm K}$ for an $s = \frac{1}{2}$ Kondo state with a characteristic temperature $T_{\rm K}$.
In URhIn$_5$, $\gamma = 50$ mJ/K$^2$mol corresponds to $T_{\rm K} = 200$ K in agreement with the characteristic temperature scale $\sim$ 150 K, suggesting the present picture is applicable.
The enhancement of $\gamma$ is consistent with the electron-electron scattering observed in electrical resistivity with the Kadowaki-Woods relation.
It should be noted, however, that the ground state of URhIn$_5$ is antiferromagnetic state occurring at 98 K which is not reflected in the discussion above.

The Fermi liquid contribution to the magnetic susceptibility, namely the Pauli susceptibility is estimated by using a Wilson ratio of 1 $\times 10^{-3}$ emu/mol at low temperature.
The experimentally obtained low temperature susceptibility at low temperature is about 1 $\times 10^{-3}$.  
Assuming an antiferromagnetic structure with ordered moments pointing in the [001] direction, the low temperature susceptibility for the [001] should vanish at $T = 0$, while the susceptibility for [100] remains constant.  
The remaining susceptibility at $T$ = 0 for $H \parallel $[001] is regarded as the contribution from conduction electrons.
Taking the Pauli susceptibility mentioned above into account, the experimental susceptibility of URhIn$_5$ can be successfully explained.

%
Another important characteristics of the magnetic susceptibility is unusually large paramagnetic Curie temperature $\theta_p$.
Based on the correction discussed above, $\theta_p$ for the [001] direction is estimated as $- 400$ K.
This value is considerably larger than the antiferromagnetic transition temperature 98 K or Kondo characteristic temperature 150 K, where both interactions are reflected in $\theta_p$.
As a result of the large $\theta_p$, the susceptibility is less temperature dependent  than that of a typical localized magnet.
This might be attributed to a partly itinerant nature of URhIn$_5$.

Finally, we compare URhIn$_5$ to UIn$_3$.
UIn$_3$ is an antiferromagnet with $T_{\rm N}$ = 88 K.  A  Kondo effect behavior seen in the temperature dependence of resistivity and susceptibility with  $A$ coefficient = 0.013 $\mu\Omega$cm/K$^2$  and $\gamma$ = 40 mJ/K$^2$mol is similar to that seen in URhIn$_5$ with a slightly different $T_{\rm K}$. The Curie Weiss behavior with $\mu_{\rm eff}$ = 3.25 $\mu_{\rm B}$/U and NMR measurements suggested localized 5$f$ behavior at high temperatures.\cite{Tokiwa01c,Sakai09} 
Hydrostatic pressure applied to UIn$_3$ increases the transition temperature : $T_{\rm N}$ = 88 K at ambient pressure monotonically increases with increasing pressure.\cite{Haga02}  On the other hand, the unit cell volume of the 'UIn$_3$' block in URhIn$_5$ is about 2 \% smaller than that of UIn$_3$ at ambient pressure, corresponding to a chemical pressure of about 2 GPa.\cite{lebihan95}  It is interesting to note that  URhIn$_5$ and UIn$_3$ under 2 GPa both have the same $T_{\rm N}$ = 98 K.  
These observations suggest that the physical properties are mainly dominated by the 'UIn$_3$' block and that the RhIn$_2$ layers are less important.

It is further interesting to compare URhIn$_5$ with URhGa$_5$.
Naively thinking one might expect a nonmagnetic ground state for URhIn$_5$ in the high pressure limit similar to URhGa$_5$, but our preliminary high pressure resistivity measurement up to 2 GPa and our thermal expansion experiment demonstrated an increase of transition temperature at high pressure.
For URhGa$_5$, dHvA experiments and band calculations have shown that an itinerant picture for its 5$f$ electrons is well established.
It will be interesting to investigate the Fermi surface in URhIn$_5$ to clarify the 5$f$ electronic state and the appearance of magnetism.
A dHvA signal has been detected in URhIn$_5$ and detailed measurements are in progress.

The volume-driven electronic change arising from valence change is commonly observed in condensed matter physics.
In particular, such changes and fluctuations can often be easily achieved in the case of strongly correlated $f$-electron systems and are widely discussed as a source for such strongly correlated phenomena as unconventional superconductivity.
The present example demonstrates that 5$f$ magnetism can appear while keeping the number of valence electrons constant at ambient pressure and  contributes to the discussion of electronic states in 5$f$ metallic systems both experimentally and theoretically.

\section{Summary}

We have prepared a new uranium compound URhIn$_5$ as high quality single crystals. 
URhIn$_5$ is an itinerant antiferromagnet with $T_{\rm N}$ = 98 K with a weakly mass enhanced  $\gamma$ = 50 mJ/K$^2$mol. 
The physical properties of URhIn$_5$ are very similar to that of UIn$_3$.

\begin{acknowledgments}
%

We thank Drs. D. Aoki, E.D. Bauer, S. Kambe, K. Kaneko, N. Metoki, Y. {\=O}nuki and Y. Tokunaga for discussions.
   This work was supported by a Grant-in-Aid for Scientific Research on Innovative Areas "Heavy Electrons (No. 20102002, No. 23102726), S (No 20224015), C (No. 22540378, 25400386) and Young Scientists B (No 24740248) from the Ministry of Education, Culture, Sports, Science and Technology (MEXT) and Japan Society of the Promotion of Science (JSPS).

\end{acknowledgments}


\end{document}